\documentclass[a4paper]{article}
\usepackage{graphicx}

\begin{document}

\begin{center}
{\Large \bf How to Improve UrQMD Model to Describe NA61/SHINE Experimental Data }
\end{center}

\begin{center}
{V. Uzhinsky}
\end{center}

\begin{center}
{CERN, Geneva, Switzerland and LIT, JINR, Dubna, Russia}
\end{center}

\begin{center}
\begin{minipage}{12cm}
The NA61/SHINE collaboration measured inclusive cross sections of $\pi^+$ and $\pi^-$
meson production in the interactions of 31 GeV/c proton with carbon nuclei at small
emission angles (0 -- 420 {\it mrad}). The collaboration presented also predictions
of Monte Carlo models -- FLUKA, VENUS and UrQMD, in a comparison with the data. The
worst description of the data was observed for UrQMD model results.

~~~~~~~~ In the present paper it is shown  that the drawback of the UrQMD model is
connected with an inaccurate treatment of low mass string fragmentation. The strings
appeared at a diffraction of target nucleons. A simple patch is proposed to overcome
the problem.
\end{minipage}
\end{center}

The discrepancy saying above between the NA61/SHINE data and UrQMD model predictions
is shown in Fig. 1. As seen, the model overestimates the experimental data  for low momenta
($p<$ 1.5 {\it GeV/c}) very strongly. Because the excess is observed for low energy mesons, it
can be supposed that a mistake is connected with a particle cascading inside the nucleus, or with
the low mass diffraction of the target nucleons.
\begin{figure}[cbth]
\includegraphics[width=160mm,height=60mm,clip]{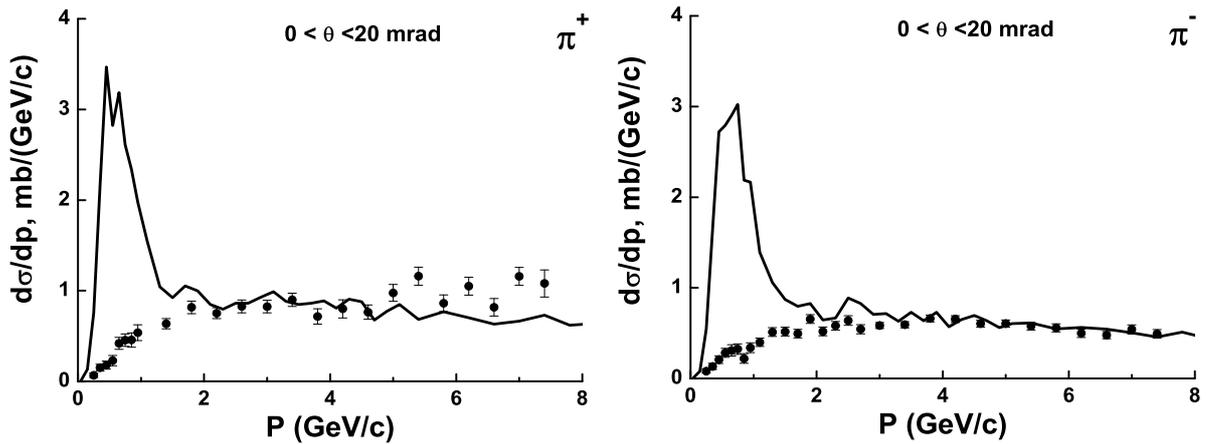}
\caption{Inclusive cross sections of meson production at $\theta =$ 0 -- 20 {\it mrad}.
Points are experimental data \protect{\cite{NA61}}. Lines are the UrQMD model 1.3
\protect{\cite{UrQMD}} predictions.}
\label{Xtotel}
\end{figure}

It was checked that a change of the cascade parameters does not affect on the predicted meson spectra.
The following calculations have been done:
\begin{enumerate}
\item The parameter {\it XAP} - string tension, in the file {\it string.f} of the UrQMD 1.3 code
      was changed from 1 to 0.1 ({\it fm/GeV}) in order to increase the particle formation time.
      One line was added in {\it subroutine leadhad} in the file {\it scatter.f} to switch out
      the leading particle effect.
      \begin{verbatim}
      subroutine leadhad(n1l,n2l,nbl)
      implicit none
      include 'newpart.inc'
      include 'comres.inc'
      include 'coms.inc'
      integer n1l,n2l,nbl,ll

c default: no leading hadron at all
       do 1 ll=n1l,n2l
         leadfac(ll)=0.d0
 1    continue
      Return                     ! Uzhi
      \end{verbatim}

\item The average transverse momentum, $P_T=1.6$ ({\it GeV/c}), transferred between interacting
      nucleons was changed to 0.4 ({\it GeV/c}) ({\it CTParam(31)=0.4}, see "User Guide") to increase produced
      particles momenta.

\item The parameter {\it Beta} was put to zero ({\it CTParam(40)=0}) to change string mass distribution.

\item {\it CTParam(2)} was put to 0.26 in order to decrease the single diffraction probability.
\end{enumerate}
{\bf All was vain!}

An analysis of a simulation of the single diffraction in $NN$-interactions shows that most of
the created strings have low masses, $\sim $ 1.3 - 1.6 ({\it GeV}). Thus, instead of a fragmentation,
a two-particles decay of a string is simulated in the {\it SUBROUTINE CLUSTR} (see the file
{\it string.f}) where the following lines are:
\begin{verbatim}
c..forward/backward distribution in clustr for baryons
c..(no pt in the last string break!)
c..pt for the baryon comes from parton kick in the excitation
      if(abs(ident(i)).ge.1000.or.abs(ident(i-1)).ge.1000)then
      PPTCL(1,I-1)=0.d0
      PPTCL(1,I)=0.d0
      PPTCL(2,I-1)=0.d0
      PPTCL(2,I)=0.d0
      PPTCL(3,I-1)=PA
      PPTCL(3,I)=-PA
      PA2=PA**2
      PPTCL(4,I-1)=SQRT(PA2+PPTCL(5,I-1)**2)
      PPTCL(4,I)=SQRT(PA2+PPTCL(5,I)**2)
      IDCAY(I-1)=0
      IDCAY(I)=0
      NPTCL=I
      endif
\end{verbatim}
As a result, the produced mesons have zero $P_T$, and they fill the region of small $p$ in Fig. 1.

There is a part for a simulation of the isotropic two-particle decay just before the lines. Thus,
it is very easy to improve the UrQMD code closing the presented lines :
\begin{verbatim}
c..forward/backward distribution in clustr for baryons
c..(no pt in the last string break!)
c..pt for the baryon comes from parton kick in the excitation
cUzhi if(abs(ident(i)).ge.1000.or.abs(ident(i-1)).ge.1000)then
cUzhi PPTCL(1,I-1)=0.d0
cUzhi PPTCL(1,I)=0.d0
cUzhi PPTCL(2,I-1)=0.d0
cUzhi PPTCL(2,I)=0.d0
cUzhi PPTCL(3,I-1)=PA
cUzhi PPTCL(3,I)=-PA
cUzhi PA2=PA**2
cUzhi PPTCL(4,I-1)=SQRT(PA2+PPTCL(5,I-1)**2)
cUzhi PPTCL(4,I)=SQRT(PA2+PPTCL(5,I)**2)
cUzhi IDCAY(I-1)=0
cUzhi IDCAY(I)=0
cUzhi NPTCL=I
cUzhi endif
\end{verbatim}

The calculation results obtained with the proposed changes are shown in Fig. 2.

\begin{figure}[cbth]
\includegraphics[width=160mm,height=60mm,clip]{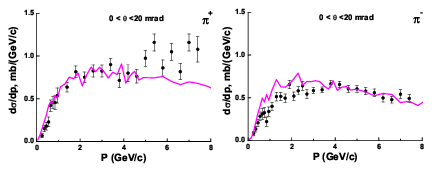}
\caption{Inclusive cross sections of meson production at $\theta =$ 0 -- 20 {\it mrad}.
Points are experimental data \protect{\cite{NA61}}. Lines are the bug fixed UrQMD model predictions.}
\label{Xtotel}
\end{figure}

~~

~~

~~

~~

The author is thankful to G. Folger and V. Ivanchenko for interest to the work and important remarks.

\end{document}